\documentclass[letterpaper,twocolumn,10pt]{article}
\usepackage{usenix2019}
\usepackage{ifthen}
\usepackage{xspace}
\usepackage{color}
\usepackage{graphicx}
\usepackage[nocompress]{cite}
\usepackage{booktabs}
\usepackage{textcomp}

\newcommand{\ANONAUTHORS}{People}

\newcommand{\TITLE}{\Large \bf EnclaveDom: Privilege Separation for Large-TCB Applications in Trusted Execution Environments}

\newcommand{\KEYWORDS}{}
\newcommand{\CONFERENCE}{} 
\newcommand{\COLOR}{no}
\newcommand{\PAGENUMBERS}{yes}

\newcommand{\ANONYMOUS}{no}

\newcommand{\COMMENTS}{yes}

\ifthenelse{\equal{\ANONYMOUS}{no}}{%
\newcommand{\AUTHORS}{\rm{\REALAUTHORS}}
\newcommand{\griffin}{EnclaveDom\xspace}
\newcommand{\enclavedom}{EnclaveDom\xspace}
\newcommand{\intelreg}{Intel\textsuperscript{\textregistered}\xspace}
\newcommand{\sgx}{Intel SGX\xspace}
}{
\newcommand{\AUTHORS}{\ANONAUTHORS}
\newcommand{\griffin}{EnclaveDom\xspace}
\newcommand{\intelreg}{Intel\textsuperscript{\textregistered}\xspace}
\newcommand{\sgx}{Intel SGX\xspace}
}

\newcommand{\eg}{e.g.,\xspace}
\newcommand{\ie}{i.e.,\xspace}

\newcommand{\Parabreak}{1.5ex}
\newcommand{\Paragraph}[1]{\vspace{\Parabreak}\noindent\textbf{#1}}

\ifthenelse{\equal{\COMMENTS}{yes}}{%
\newcommand{\msm}[1]{\textbf{MSM: #1}}
} {
\newcommand{\msm}[1]{}
}

\usepackage{fancyhdr} 

\ifthenelse{\equal{\PAGENUMBERS}{yes}}{%
  \pagestyle{plain}
}{%
  \pagestyle{empty}
}

\date{}
\title{\TITLE}
\author{{\AUTHORS}%
\ifthenelse{\equal{\ANONYMOUS}{no}}{%
  \\*[0.1in] {$^1$Intel Labs, $^2$Princeton University}}{}
}

\usepackage[hyphens]{url}
\ifthenelse{\equal{\COLOR}{yes}}{%
  \usepackage[colorlinks]{hyperref}
  \definecolor{darkblue}{RGB}{0,0,120}
  \definecolor{reallydarkblue}{RGB}{0,0,63}
  \definecolor{Black}{RGB}{0,0,0}
  \hypersetup{
    colorlinks,
    citecolor=Black,
    linkcolor=reallydarkblue,
    urlcolor=darkblue}
}{%
  \usepackage[pdfborder={0 0 0}]{hyperref}
}

\hypersetup{
pdfauthor = {\AUTHORS},
pdftitle = {\TITLE},
pdfsubject = {\CONFERENCE},
pdfkeywords = {\KEYWORDS},
bookmarksopen = {true}
}

\begin{document}

\maketitle

\ifthenelse{\equal{\PAGENUMBERS}{yes}}{%
}{\thispagestyle{empty}}

\begin{abstract}
Trusted executions environments (TEEs) such as \intelreg SGX
provide hardware-isolated execution areas in memory,
called \emph{enclaves}. By running only the most trusted application
components in the enclave, TEEs enable developers
to minimize the TCB of their applications thereby
helping to protect
sensitive application data.
However, porting existing applications to TEEs often
requires considerable refactoring efforts,
as TEEs provide a restricted
interface to standard OS features.
To ease development efforts, TEE application
developers often choose to run their unmodified
application in a library OS container that provides
a full in-enclave OS interface.
Yet, this large-TCB development approach now
leaves sensitive in-enclave data exposed to potential
bugs or vulnerabilities in \emph{third-party} code imported
into the application.
Importantly, because the TEE libOS and the application
run in the same enclave address space, even the
libOS management data structures (e.g. file descriptor table)
may be vulnerable to attack,
where in traditional OSes these data structures
may be protected via privilege isolation.

We present \griffin, a privilege separation
system for large-TCB TEE applications that
partitions an enclave into tagged memory regions, and enforces
per-region access rules at the granularity of
individual in-enclave functions.
\griffin is implemented on \sgx using Memory
Protection Keys (MPK)~\cite{intel-sdm} for memory tagging.
To evaluate the security and performance impact
of \griffin, we integrated \griffin with the
Graphene-SGX~\cite{graphene-sgx} library OS.
While no product or component can be absolutely secure,
our prototype helps protect
internal libOS management data 
structures against tampering by application-level code.
At every libOS system call, \griffin then only grants access to 
those internal data structures which the syscall 
needs to perform its task.
Our \griffin prototype imposes reasonable performance and
modest memory overheads on Graphene-SGX.
\end{abstract}

\section{Introduction}

Trusted execution environments (TEEs) such as
\sgx~\cite{sgx, sgx-paper}
enable developers to create execution areas,
called \emph{enclaves} with enhanced protection
within a CPU.
At its core, a TEE aims to provide confidentiality
of sensitive application code and data in the presence of
untrusted or vulnerable system software.
In other words, the goal of TEEs is to allow developers to
reduce the TCB of applications
by placing only the most security-critical functionality 
in the enclave, and leaving the majority of
application function the \emph{untrusted} context.

Yet, deploying existing applications to TEEs typically requires
that developers significantly re-architect
their applications. In the case of \sgx, for instance,
applications may not make direct system calls from within an enclave.
Thus, \sgx application development has been shifting towards a
containerized model, in which unmodified applications run on top of
a TEE-specific library OS (libOS) inside an 
enclave~\cite{graphene-sgx, panoply, scone, haven, sgxkernel}.
These libOSes provide a general system call interface
that transparently handles all enclave-to-untrusted transitions
needed to call system software.

However, this practice vastly increases the TCB of an
application, which now includes
any untrusted third-party code as well as the underlying TEE libOS.
Including third-party libraries within an enclave is problematic
for two main reasons. First, application programmers
rarely have the time, expertise, or even authority,
to prioritize security and privacy in their development
process~\cite{balebako,acar1,acar2}.
Thus, developers cannot be expected to fully inspect
the source code of every library they need for their application,
meaning that any vulnerabilities or bugs in imported third-party software
may remain undetected.
Indeed, software supply chain attacks are becoming increasingly
common, as adversaries leverage the widespread use of
open-source libraries to disseminate malicious
code (\eg~\cite{pypi-attack,npm-attack}).

Nevertheless, the convenience of 
third-party libraries makes them indispensable to today's
software development practices.
Developers rely on community trust,
especially in open-source libraries, hoping that
the third party whose code they import has not published malicious
code and otherwise done their due diligence to
eliminate any security vulnerabilities.
For example, in spite of numerous 
examples of data leak bugs in security-critical libraries such as 
OpenSSL~\cite{openssl-vulns}, developers still opt to
include such libraries in their TEE application.

Second, by including unvetted third-party code in their TEE
application, developers are no longer using TEEs for their originally
intended purpose:
TEEs are designed to run a small set of trusted application
components, which do not need to be run with different privileges.
In a large-TCB application, in which different components may
require different access privileges to sensitive data,
all code running inside an enclave
still runs with the same privileges.
As such, all third-party code imported into an application running inside
a TEE libOS container has unfettered access to all enclave memory
leaving the application susceptible to data leaks and corruption.

To make matters worse, most TEE libOSes proposed thus far do not enforce
privilege isolation. This crucial security mechanism in traditional OSes
distinguishes between user-level and kernel-level processes,
ensuring that userspace processes cannot access kernel-level memory. 
On the other hand, TEE libOSes execute in userspace
alongside the application they are running. 
So, despite running inside a TEE, 
the libOS's internal management data structures,
such as the file descriptor table or the mount table, 
may be corrupted by vulnerable or malicious third-party
code imported by the \emph{application} developer.

To enable developers to reap the benefits of TEEs
while making their applications more robust in the face of
unvetted third-party libraries, we present \enclavedom, an 
in-enclave privilege separation system for large-TCB
applications running inside TEEs.

Privilege separation in legacy applications has been the subject
of a large body of prior proposals, all with the common
goal of enforcing \emph{least privilege}~\cite{least-priv}.
Similar to approaches that leverage process isolation to
restrict different application components'
access to sensitive data or OS resources
(\eg~\cite{passe,breakapp,privtrans}), prior
research addressing privilege separation
in \sgx applications, for instance,
has proposed running individual application components in
multiple separate \sgx processes
with communicating enclaves~\cite{graphene-sgx, panoply, ryoan}.

Yet, these systems either only control access to
sensitive data at a per-enclave granularity~\cite{graphene-sgx, ryoan},
or they do not protect the TEE libOS or shim layer itself
against application code~\cite{graphene-sgx, panoply}. 
Furthermore, this approach
requires inter-enclave communication between the 
different processes to compute on shared data,
which incurs significant performance overheads.

In contrast, \enclavedom subdivides a \emph{single} enclave 
into multiple isolated memory compartments, each with its own access policy.
\enclavedom then stores developer-specified sensitive
in-enclave data objects, such as TLS keys or sensitive datasets, 
in these compartments. Thus, \enclavedom can enforce fine-grained
least privilege at the granularity of \emph{in-enclave functions}
without refactoring the application
into multiple TEE processes.

One major challenge in \enclavedom's single-enclave design
is to securely share sensitive enclave data between
functions with different access privileges to the same data.
\enclavedom addresses this issue using hardware-assisted
memory tagging.
Thus, \enclavedom creates different \emph{memory domains} 
within an enclave by assigning memory tags to enclave pages.

While the mechanisms we present in this paper apply
generally to TEEs, we realize \enclavedom on top of \sgx using
Memory Protection Keys (MPK),
a memory tagging technique developed at Intel.
MPK provides a special hardware register
that stores a process' access privileges to each MPK tag, and
enforces two types of access privileges to tagged pages 
(read-only and read-write). 
Userspace processes may then dynamically adjust the 
access privileges for different memory tags.

\enclavedom targets two \sgx deployment settings: (1) containerized
applications running in a TEE libOS, and (2) \sgx-native applications
developed using the \sgx SDK~\cite{sgx-sdk}.
Although \sgx-native applications typically have
a smaller TCB than the libOS setting, developers often still incorporate
unvetted third-party libraries into their enclave code for ease of development.
Thus, in both settings, \enclavedom helps developers enhance the protection of
sensitive in-enclave data against leaks or tampering by third-party libraries
running in the enclave. In the TEE-libOS setting, \enclavedom can
additionally be used to implement OS privilege isolation.

To declare sensitive in-enclave data objects and define those
in-enclave functions that are authorized to access them,
developers in \enclavedom specify the sensitive input arguments
or return values of third-party library functions executing
inside in the enclave in a central policy file. At run time, \enclavedom
then assigns each developer-specified data object to
an MPK memory domain, and only grants access to those domains when
a privileged in-enclave function is executing.

\enclavedom does not require extensive
manual annotations to application source code by the developer, and
can be integrated into TEE libOSes to transparently
improve the protection of
the containerized applications they run.
We demonstrate the effectiveness of our approach by porting \enclavedom
to the Graphene-SGX libOS~\cite{graphene-sgx}. As a preliminary
step, our prototype implements privilege isolation in Graphene-SGX to
help protect the internal libOS data structures, including the
file descriptor and mount point tables, against leaks and tampering
by untrusted application-level code running on top of the libOS in
the enclave. 

Although we needed to modify the Graphene-SGX source code
to add support for \enclavedom,
applications still run completely unmodified inside the TEE libOS container. 
To analyze the security properties of \enclavedom,
we first study an adversarial application that
mounts a libOS data corruption attack
in our Graphene-SGX prototype, and then
examine \enclavedom's ability to help
prevent additional
hypothetical data leak and corruption attacks mounted
within enclave code.
We evaluated our \enclavedom prototype's performance via
Graphene-SGX system call microbenchmarks, and find that
\enclavedom imposes acceptable performance and modest
memory overheads.

\section{Background}

\begin{figure}[t]
  \centering
  \includegraphics[width=0.47\textwidth]{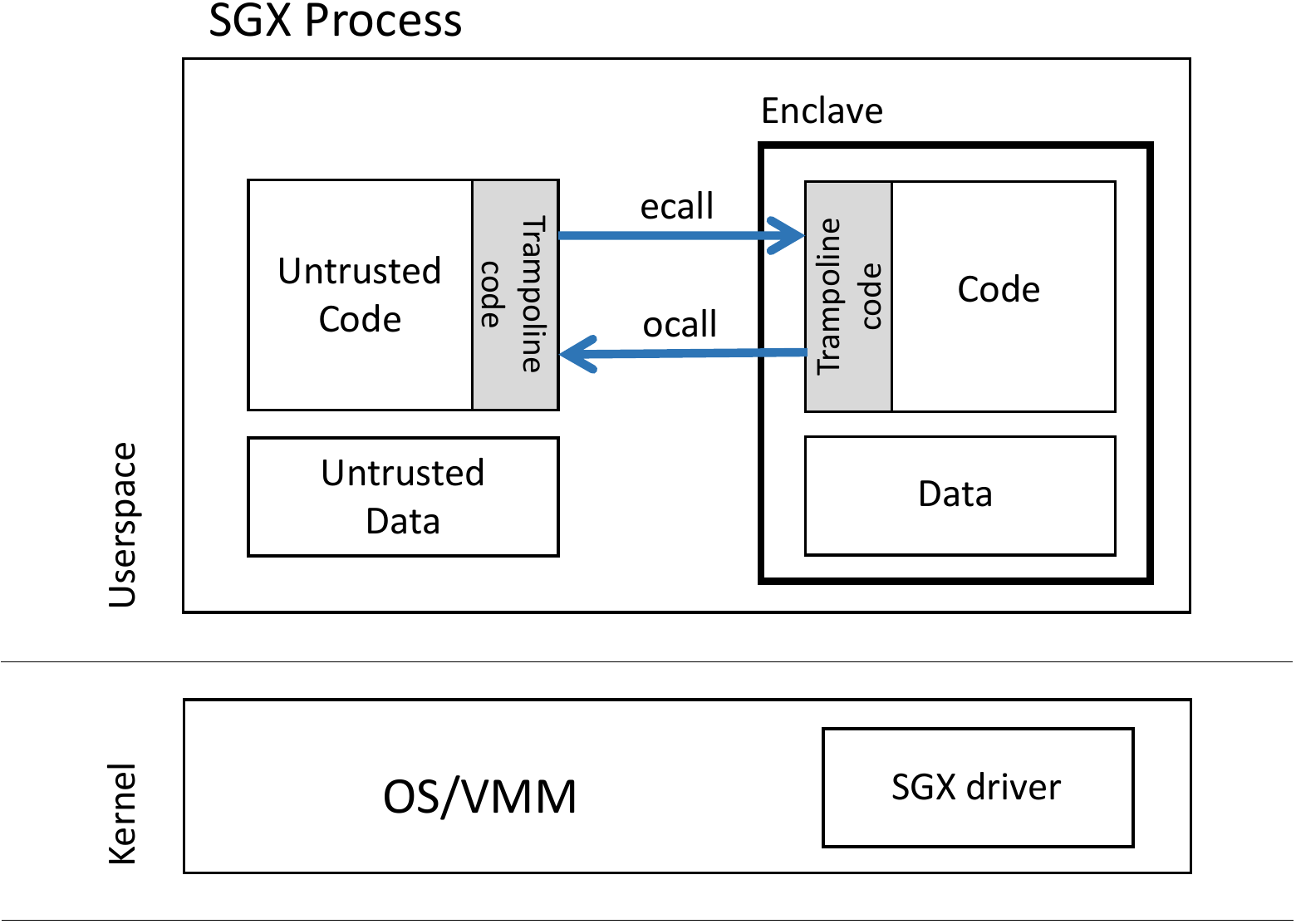}
    \caption{\label{fig:sgx-mem} \intelreg SGX application memory layout.}
\end{figure}

We instantiate the \enclavedom memory access control system as an experimental
combination of \sgx and Memory Protection Keys.
This section describes these two techniques, and summarizes 
the programming framework for \sgx applications.

\subsection{\intelreg SGX}
\label{sec:sgx-background}
Intel Software Guard Extensions (SGX)~\cite{sgx,sgx-paper} 
is a trusted execution environment technology
that is designed to preserve the confidentiality
of application code and data, even in the face of
untrusted or compromised system software.

At its core, \sgx provides an isolated memory region
within the address space of a userspace process, called
an \emph{enclave}. All code and data within the enclave is encrypted
for the entire lifetime of the application.
This design divides an application into a trusted
and an untrusted context, where the trusted context in the enclave
is isolated even from the underlying OS or VMM.

\sgx dictates that the enclave must be entered 
via pre-defined entry points called \emph{ecalls}, 
and requires that all system calls be made through 
special trampoline functions that exit the enclave, called \emph{ocalls}.
At every ecall and ocall, \sgx encrypts or decrypts any 
memory contents being passed
between the contexts.
Further, \sgx cryptographically
computes an \emph{enclave measurement} at application 
startup, which \sgx applications may use to attest to the
authenticity of the enclave code. Fig.~\ref{fig:sgx-mem}
shows the high-level memory layout of an \sgx application.

\begin{figure}[t]
  \centering
  \includegraphics[width=0.47\textwidth]{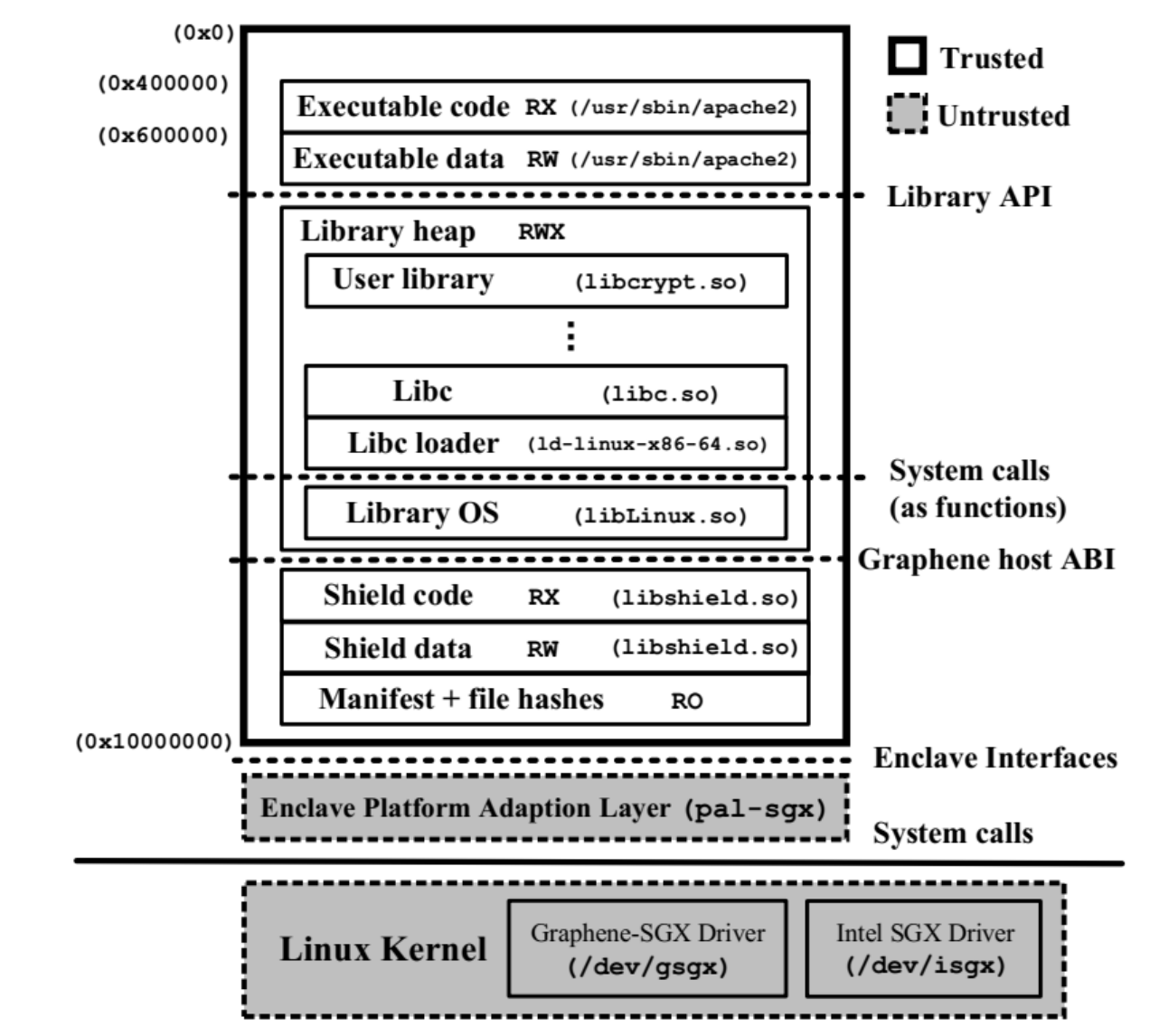}
  \caption{\label{fig:graphene-arch} The Graphene-SGX system architecture
    (copied from~\cite{graphene-sgx}).}
\end{figure}

\subsection{LibOS-based Containers for \intelreg SGX}
\label{secsec:libos-dev}

\sgx-specific library OSes (libOS) such as Graphene-SGX~\cite{graphene-sgx},
SCONE~\cite{scone}, Panoply~\cite{panoply}, and SGXKernel~\cite{sgxkernel},
are designed to facilitate the deployment of
\sgx applications by transparently handling all enclave-to-untrusted
transitions necessary to utilize standard OS features.
To this end, libOSes provide a trusted shim layer that implements
a full userspace-level system call interface essentially creating
an in-enclave container that runs unmodified applications
on top of the libOS. As an example, Fig.~\ref{fig:graphene-arch}
shows the system architecture for the Graphene-SGX libOS.

We present a prototype of \enclavedom based on Graphene-SGX,
but emphasize that our mechanisms are not specific to Graphene-SGX.
One key feature of Graphene-SGX is its support of dynamically loaded code,
requiring only very few additional steps to develop an application
for Graphene-SGX.
To deploy an application, developers
must first specify the resources (\eg files and network rules)
required by the application in an application-specific manifest file.
Following \sgx requirements, the manifest also specifies certain
enclave parameters, such as the maximum enclave size and maximum number
of threads, which developers may configure based on their application's
needs. Developers may then run a Linux executable via
the Graphene-SGX command-line tool.

\subsection{\intelreg SGX-native Software Development}
\label{secsec:sgx-dev}

In order to provide \sgx's security properties,
\sgx requires that programmers follow two main development rules
when creating an \sgx-native application.

First, all enclave code subject to enclave measurement must be static,
\ie built as a statically
linked library as part of the whole application. In other words, a major
task facing \sgx application developers is defining a priori all of the code
that runs within the enclave, as well as the ecalls and ocalls
for transitioning into and out of the enclave.

To alleviate this effort,
the \sgx SDK~\cite{sgx-sdk} provides the \emph{Edger8r} tool
which automatically generates ecall and ocall trampoline functions based
on a developer-specified enclave definition file specifying the function
signatures for the functions implementing the enclave entry and exit interface.
More specifically, Edger8r generates the source and header
files for the pre-defined trampoline code, which can then be included as part
of the enclave code.

Second, enclave code must be digitally signed to allow \sgx to verify
the enclave measurement as well as verify the legitimacy of the enclave
author. To this end, the \sgx SDK also provides the \emph{Sign Tool} that
automatically computes the enclave measurement and corresponding
signature given the enclave code and the application developer's enclave
signing key. Upon enclave startup, \sgx can then verify the measurement
and signature to confirm the authenticity of the application's enclave.

\subsection{Memory Protection Keys (MPK)}
\label{sec:mpk-background}
Memory Protection Keys (MPK)~\cite{intel-sdm} are a recently-developed
feature of the Intel x86 architecture,
which enables userspace processes to tag each page table entry
with a 4-bit \emph{protection key}.
Thus, applications create up to 16 isolated \emph{memory domains}
within their process address space, and assign individual pages to
different domains.

MPK introduces the PKRU, a per-core 32-bit register that stores a bitmap of read-only and read-write access
bits for the 16 protection keys. Whenever a process requests access to a page,
the memory management unit checks the PKRU to verify that the process has sufficient privileges to
access the requested address. To adjust the access permissions to a
given domain,
an application calls the \texttt{WRPKRU} instruction.
Though not available on any publicly released CPUs
at the time of writing, MPK may also be used to tag \sgx enclave pages,
providing an additional layer of security to \sgx enclave memory. 

\section{System Model and Design Goals}
\label{sec:sys-model}

The goal of \enclavedom is to provide a privilege separation system that helps
protect large-TCB applications running inside a TEE, \ie
containerized applications running inside a TEE libOS as well
as large TEE-native applications, against
data leaks and corruption by in-enclave third-party code.
\enclavedom combines the security properties of TEEs and
hardware-enforced memory tagging to isolate sensitive in-enclave data objects,
and only allow privileged in-enclave functions to access these data objects.

\subsection{Threat Model}
\label{sec:threat-model}
As in TEEs, \enclavedom does not trust any hardware
outside the CPU, system software including the
OS and any hypervisor, other applications running
on the same machine alongside the TEE application,
as well as the untrusted component of the application
running outside the enclave.

However, \enclavedom assumes a large TCB programming model
for TEEs. That is, application developers wish to deploy
TEE-protected applications while incorporating third-party
libraries that have not been fully vetted (if at all).
Thus, while the TEE application developer may have
good intentions, \enclavedom does not trust third-party code
the developer imports into her application as part of the enclave code,
since it may contain undiscovered data leak vulnerabilities
or attacks. 
Nevertheless, \enclavedom maintains trust in the TEE libOS container
running unmodified applications, as well as in the \sgx SDK and its tools
used to implement \sgx-native applications.

Importantly, \enclavedom does not aim to mitigate control
flow attacks such as buffer overflow or ROP attacks. Such
attacks are typically performed as a pathway for an adversary to
gain control over the application and execute attack code.
Yet, because the adversary's
code has already inadvertently been included in the enclave
via a third-party library, \enclavedom assumes that this is sufficient
to allow the attacker to moderate the control flow of the enclave.
Thus, \enclavedom's primary
goal is to prevent \emph{unauthorized data accesses} that lead to
data leaks, corruption or enable other security attacks.
Techniques complementary to our approach (\eg~\cite{sgxbounds,intel-cet})
may be employed to safeguard applications against control flow attacks.

\enclavedom also does not address side channel or enclave API misuse
vulnerabilities such as Iago attacks~\cite{iago,ryoan};
countermeasures to these vulnerabilities have been 
proposed in prior research (\eg~\cite{ryoan,glamdring,tsgx,varys,cloak})
and are complementary to our work. 
Much like prior proposals, we also do not address denial-of-service attacks. 

\subsection{Security Properties}
Our design for \enclavedom provides the following three
security properties.

\Paragraph{P1: Least Privilege.} An in-enclave function may only access those sensitive enclave data 
objects that this function needs to provide its expected functionality. \enclavedom achieves such fine-grained least
privilege by placing developer-specified sensitive enclave data objects in isolated memory compartments,
and enforcing a \emph{default-deny} access policy to these compartments at the granularity of individual
functions running inside the enclave. 

\Paragraph{P2: Single Enclave Isolation.} Strong isolation of sensitive in-enclave data does not
require partitioning a TEE application into multiple enclaves. \enclavedom relies on a hardware-assisted
memory isolation technique capable of controlling access to individual enclave pages,
enabling the creation of isolated memory compartments \emph{within} a single enclave.

\Paragraph{P3: Secure Data Sharing.} Multiple in-enclave functions may need to
operate on the same sensitive data object. To enable data sharing between functions
that may have different access privileges within a single enclave, \enclavedom's memory compartments
support dynamic access privileges. That is, \enclavedom enforces access permissions to a given
sensitive data object based on the function requesting access.

\Paragraph{Non-goals.}
\enclavedom automatically controls access to sensitive 
data inside a TEE at the level of in-enclave functions.
However, \enclavedom does not aim to provide automated application
code partitioning as in Glamdring~\cite{glamdring}. 
Such execution isolation is orthogonal to our approach,
and could be used in conjunction with \enclavedom to further reduce
the TCB of TEE applications.
We assume that the 
application developer has partitioned her application into trusted
and untrusted components either manually or using an
automated technique.

\enclavedom also does not ensure the \emph{correctness}
of the sensitive in-enclave data it helps
protect;
While providing these stronger security
features may improve
the integrity of sensitive enclave inputs and help prevent
 additional data leaks via buggy or malicious third-party code at run time,
formally verifying the implementation of enclave code, as well as the provenance of enclave inputs,
 is orthogonal to \enclavedom's goals.

\section{\enclavedom Design}
\label{sec:design}

Our design for \enclavedom provides isolation and access
control for sensitive in-enclave data objects in the face
of untrusted third-party enclave code, without partitioning
the TEE application into multiple enclaves. In~\S\ref{sec:implementation},
we describe how \enclavedom does not require extensive
manual annotations to application source code by the developer,
and how libOSes can integrate \enclavedom to transparently
improve the protection of
the containerized applications they run.

Much as in prior research~\cite{pyronia,flexdroid}, \enclavedom relies on
the application developer's understanding of the libraries that they
import, and on their high-level expectations of how these libraries
access sensitive application data. Based on this information,
\enclavedom requires developers to specify access privileges for individual
in-enclave functions to specific sensitive data objects.

\begin{figure}[t]
  \centering
  \includegraphics[width=0.47\textwidth]{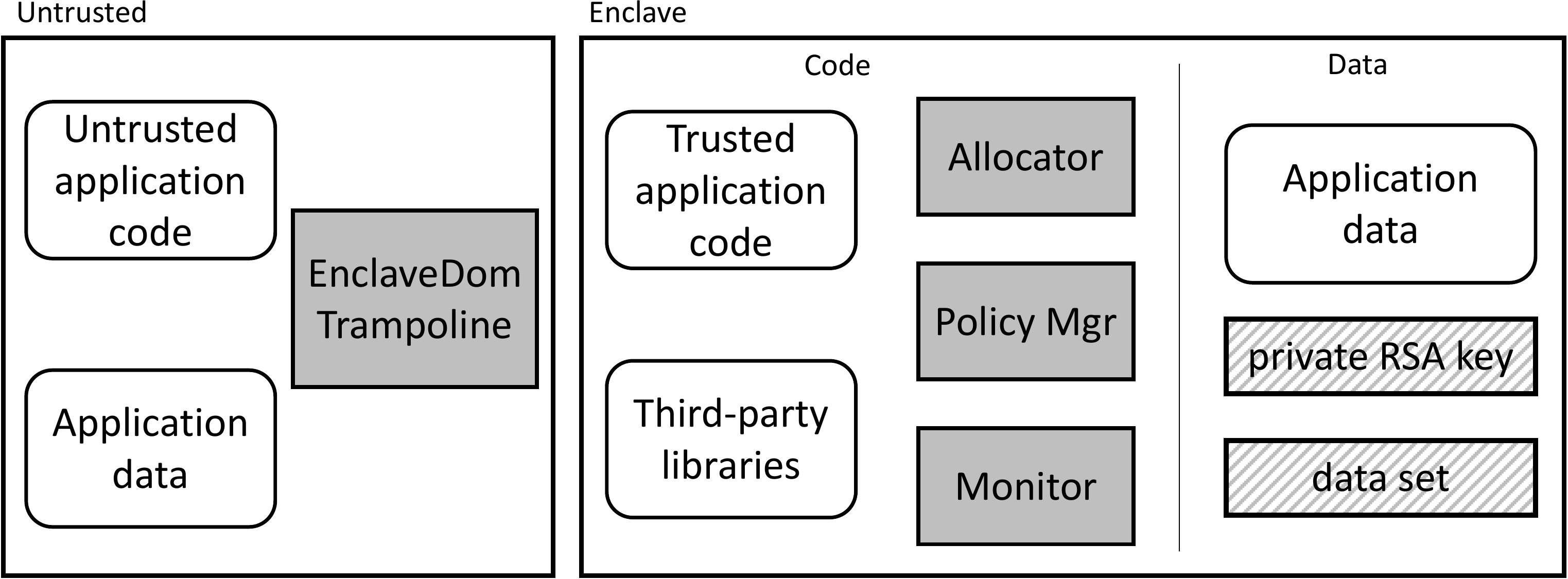}
  \caption{\label{fig:enclavedom-arch} Overview of \enclavedom, which isolates
    sensitive in-enclave data in memory domains (striped boxes),
    and controls access to these domains with three key
    components (gray boxes).}
\end{figure}

At enclave startup,
\enclavedom maps these data objects to in-enclave memory
domains~\S\ref{secsec:memdoms} and partitions the enclave into
corresponding compartments based on the developer's
policy~\S\ref{secsec:allocator}.
Then at run time, \enclavedom monitors the enclave execution, and dynamically
adjusts the enclave's access to the memory domains according to the currently
executing function~\S\ref{secsec:monitor}. Figure~\ref{fig:enclavedom-arch}
shows the main components of the \enclavedom system.

\subsection{Primitives}
\label{secsec:memdoms}
\enclavedom introduces three primitives that aim to provide usable yet strong
protection improvement of in-enclave data.

A \textbf{\emph{sensitive data object}} represents a unit of in-enclave data
that developers wish to protect. Each data object
has an associated \emph{label}, a human-readable string uniquely identifying
an object in the policy and \enclavedom run-time monitor.
Examples of sensitive data objects
might be a cryptographic key in an encrypted
data processing application such as Opaque~\cite{opaque},
or a data buffer containing a genome data set
in a genomics application.

\enclavedom uses \textbf{\emph{memory domains}} to represent isolated memory
compartments that hold one or more sensitive data objects within an enclave
address space. Each domain is then mapped to its own MPK tag to
control access to the domain at run time.
To specify in which domain \enclavedom will place a given data
object, developers may also assign a unique label to each domain in
their policy.

At run time, \enclavedom creates a \textbf{\emph{dynamic execution sandbox}},
to execute a given in-enclave function with temporarily elevated
access privileges to those memory domains specified in
the developer's policy. Upon returning
from this function, \enclavedom immediately revokes access to those
memory domains to limit any further access outside the scope of the
privileged in-enclave function.

\begin{figure*}[t]
  \centering
  \includegraphics[width=0.67\textwidth]{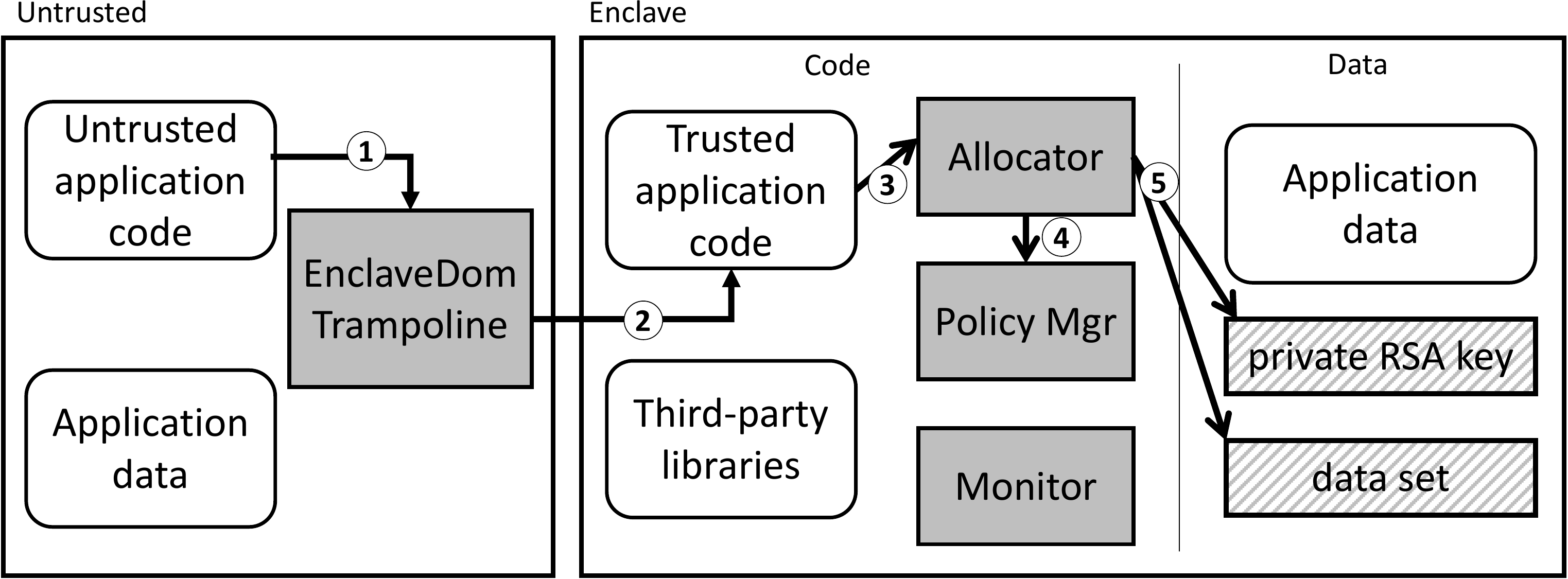}
  \caption{\label{fig:dom-alloc} The steps that \enclavedom takes
    at enclave startup to partition an enclave into memory domains.}
\end{figure*}

\begin{figure}[t]
  \centering
  \includegraphics[width=0.47\textwidth]{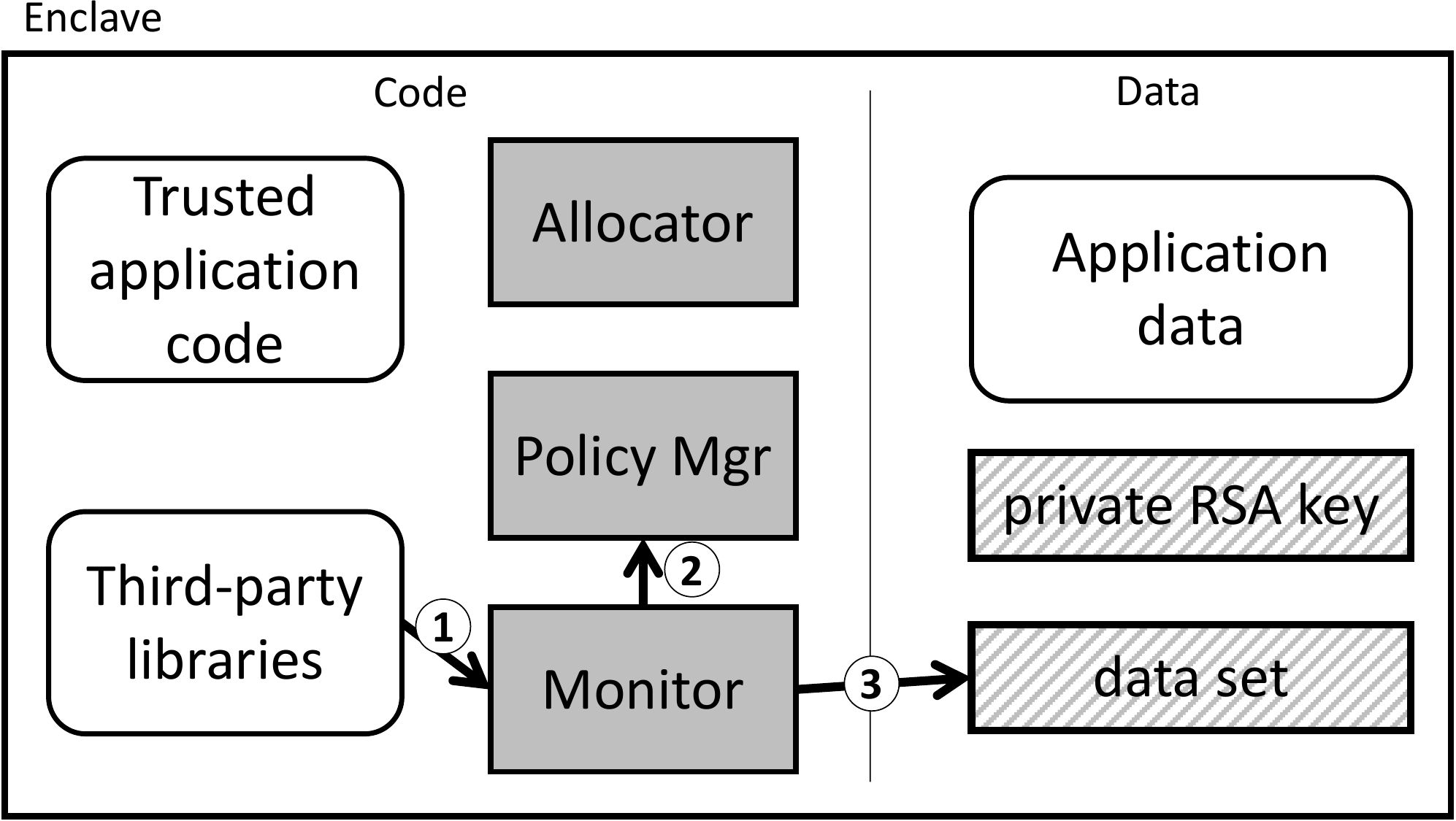}
  \caption{\label{fig:exec-monitor} The steps that \enclavedom takes
    when entering an in-enclave function that may require access to
    domain-protected data.}
\end{figure}

\subsection{Enclave Compartmentalization}
\label{secsec:allocator}

The \enclavedom Policy Manager maintains an access control list (ACL)
based on the developer-supplied data access policy.
As we show in Fig.~\ref{fig:dom-alloc}, at enclave startup,
the \enclavedom Allocator partitions the
enclave into memory domains based on the ACL, and is then responsible
for allocating sensitive data objects in their
corresponding domains at run time.

Since \enclavedom aims to improve the protection of
a wide variety of sensitive data objects,
memory domains support flexible data object sizes.
To this end, each memory domain consists of a \emph{pool} of enclave pages
all tagged with the same MPK tag.
Notably, developers need not know the exact size of their sensitive data
objects in order to isolate them in a domain since \enclavedom allows
developers to dynamically allocate domain memory at run time.

This mechanism allows \enclavedom to support arbitrary and/or variable
data object sizes. For added flexibility, developers may configure the
maximum number of pages per pool based on an
\emph{estimate} of the maximum memory consumption of their
specified data objects.

It is important to note that there is a trade-off between the granularity of
data object isolation (\ie security) and an application's
memory footprint.
Recall from ~\S\ref{sec:mpk-background} that
changes to the PKRU affect all memory pages with the same MPK tag.
This means that, on the one hand, if an application's memory domains
contain only single, small data objects,
most of the reserved enclave pages in the domains'
pools will remain unused throughout the lifecycle of the application
inflating the memory footprint of the application.

However, this policy requires that functions
have sufficient privileges to access the
memory domains corresponding to each data object.
On the other hand, a developer may use the reserved page pool more efficiently
by placing multiple data objects in the same memory
domain, at the risk of potentially allowing a function that \emph{should}
only access one of those data object to leak another sensitive data object
in the same domain.

\subsection{Dynamic Sandboxed Execution}
\label{secsec:monitor}

MPK restricts access to memory pages at the CPU core level
meaning that \emph{all} code running on the same core has access
to all tagged pages whose PKRU access bits make those pages accessible.
However, to enforce least privilege, \enclavedom cannot allow
enclave functions with insufficient privileges to access specific
data objects.

To enhance the protection of
sensitive data objects in face of leaks or tampering by
an untrusted third-party library at run time, \enclavedom helps to
ensure that only those memory domains to which the currently executing
in-enclave function has privileges are accessible, and denies
access to all other domains. Specifically, upon entering a privileged
in-enclave function, the \enclavedom Monitor queries the Policy Manager
for the given function's data object access rules, and encapsulates
the function in an execution sandbox by dynamically adjusting
the access bits for the MPK tags associated with the corresponding
memory domains (see Fig.~\ref{fig:exec-monitor}).

After returning from the sandboxed function, and before exiting
the execution sandbox,
the \enclavedom Monitor then once again restricts access to all memory domains.
While this step may be avoided if the next enclave function to be
executed also has access to some or all of the accessible domains, \enclavedom
takes a conservative approach and makes no assumptions about the control
flow of applications.

\begin{table*}[t]
  \caption{\label{tab:api} The \enclavedom API. In a Graphene-SGX container
    deployment setting, the API is used by the \emph{libOS} developers.
    For an \sgx-native deployment setting, the * denotes API calls that
    can be automatically generated based on the developer's policy.}
  \begin{center}
  \begin{tabular}{l}
    \toprule
 \textbf{Initialization/teardown} \\
 \hline
 \texttt{int enclavedom\_init(void)} \\
 \texttt{void enclavedom\_teardown(void)} \\
 \midrule
 \textbf{Data object management}* \\
 \hline
 \texttt{int enclavedom\_check\_input\_size(const char *obj\_label, size\_t size)} \\
 \texttt{int enclavedom\_check\_output\_size(const char *obj\_label, size\_t size)} \\
 \texttt{int enclavedom\_copy\_from\_untrusted(const char *obj\_label, void *untrusted\_buf)} \\
 \texttt{int enclavedom\_copy\_to\_untrusted(const char *obj\_label, void *untrusted\_buf)} \\
 \midrule
 \textbf{Domain memory management}* \\
 \hline
 \texttt{void *enclavedom\_malloc(const char *domain\_label, size\_t size)} \\
 \texttt{void enclavedom\_free(const char *domain\_label, void *addr)} \\
 \midrule
 \textbf{Execution sandboxing}* \\
 \hline
  \texttt{int enclavedom\_grant\_data\_access(const char *func\_name)} \\
  \texttt{void enclavedom\_revoke\_data\_access(const char *func\_name)} \\
\bottomrule
  \end{tabular}
  \end{center}
\end{table*}

\begin{table*}[t]
  \caption{\label{tab:access-rule-spec} Supported data access semantics
    and their corresponding policy rule specification.}
  \begin{center}
  \begin{tabular}{cc}
    \toprule
 \textbf{Access semantics} & \textbf{Object specification} \\
 \hline
 no writable objects & $O_O = \emptyset$  \\
 all writable objects & $O_I = \emptyset$ \\
 blanket access to domain & \texttt{\#<domain label>:} \\
 skip object size verification & \texttt{<object label>\#<domain label>:} \\
\bottomrule
  \end{tabular}
  \end{center}
\end{table*}

\section{Implementation}
\label{sec:implementation}

We implement \enclavedom as a userspace library that can be
integrated into \sgx-specific libOSes and
\sgx-native applications in a handful of steps,
and we demonstrate \enclavedom's practicality and benefits
with our \enclavedom-protected Graphene-SGX libOS prototype.

\subsection{\enclavedom API}
\label{secsec:enclavedom-api}

To facilitate adoption and usability, our userspace library
for \enclavedom exposes a small programming interface which abstracts away
the use of MPK.
Importantly, the execution sandboxing API calls can be automatically
generated in both of \enclavedom's targeted deployment settings.
For developers writing \sgx-native applications, we envision providing
a modified version of the \emph{Edger8r} tool to
minimize the number of manual changes to application source code.
Further, to allow developers of containerized applications (\ie those
running inside a libOS) to improve the protection of
sensitive data via \enclavedom, we could
provide a tool for libOS developers to transparently enforce the
application developer's data access policy and automate execution sandboxing.

We also provide a policy generation tool that automatically
populates header files with the internal representation of the application's
data object ACL (see~\S\ref{secsec:enclavedom-code}).
To maintain the integrity of the
developer's data access policy, we do not provide
any policy-related API calls; the internal representation of the policy
is only accessible within the \enclavedom code.
Table~\ref{tab:api} shows \enclavedom's API.

\Paragraph{Initialization and teardown.}
In order to enable \enclavedom's run-time memory access control,
developers must use the initialization and teardown API.
Notably, these functions should only be called once each, and may be called
either from the untrusted application context or from within the enclave.
Given the data object ACL generated at enclave build time,
\texttt{enclavedom\_init()} is responsible for
pre-allocating the enclave pages reserved for memory domains, and for
provisioning and mapping MPK tags to memory domains. Correspondingly,
\texttt{enclavedom\_teardown()} releases provisioned MPK tags with the OS,
and frees the memory domain enclave pages.
These two API calls are the only \enclavedom functionality that libOS developers
and \sgx-native application developers \emph{must} manually incorporate
into their application source code.

\Paragraph{Data object management.}
\sgx-native application developers may wish to pass data into
the enclave and treat it
as sensitive data, or return a data object to the untrusted context as the
result of an enclave computation.
To support passing sensitive data objects during ecalls and ocalls,
\enclavedom provides developers with an API for
copying a given sensitive data object into or out of the enclave.
As part of passing sensitive data objects between contexts, \enclavedom
also allows developers to verify the size of the data objects. This API
may only be called from within an enclave.
To ease adoption, these API calls can be easily integrated into
the \sgx SDK's \emph{Edger8r} edge routine generation tool.

\Paragraph{Domain memory management.}
As we describe in~\S\ref{secsec:allocator}, applications in \enclavedom may
allocate data in memory domains dynamically at run time. \enclavedom's memory
management API allows libOS and \sgx-native application developers to
either replace or complement existing malloc calls inside
the enclave with domain-specific memory allocations.
As with the data object management API, this functionality can be
integrated into the \emph{Edger8r} tool to facilitate the adoption process
for \sgx-native applications.

\Paragraph{Execution sandboxing.}
While developers must explicitly specify each sensitive data object that
a privileged in-enclave function may access in their policy, \enclavedom
transparently maintains all data object-to-domain mappings. Developers then
use \enclavedom's execution sandboxing API to specify the boundaries of
a function sandbox by encapsulating privileged function calls between
a \texttt{enclavedom\_grant\_data\_access()} call at the entry of a
privileged function, and the corresponding
\texttt{enclavedom\_revoke\_data\_access()} call after returning.
In~\S\ref{secsec:enclavedom-code}, we describe how \enclavedom can help to automate
this process for application developers in both deployment settings.

\subsection{Policy Specification}
\label{secsec:policy-spec}
\enclavedom requires application developers to declare sensitive data objects
and specify corresponding enclave function-level access rules in
a static policy file. In addition, libOS developers may also generate a
\enclavedom policy to improve the protection of
internal libOS data structures that is included
in the libOS core (see~\S\ref{secsec:graphene-prototype}).
Every access rule takes the form:

\texttt{$O_I$ > $func$ > $O_O$} \\
where $func$ is the name of the in-enclave function affected by the given
rule. $O_I$ specifies those sensitive data objects to which $func$ has
read-only access (\ie function inputs), and $O_O$ specifies those
data objects to which $func$ has read-write access (\eg function return
values).

In an access rule, $O_I$ and $O_O$ are written as comma-separated lists of
\emph{object specifications} formatted as

\texttt{<object label>\#<domain label>:<object size>}. \\
To allow developers to tailor their policies to their application's security
and performance requirements,
\enclavedom supports flexible data access semantics, detailed in
Table~\ref{tab:access-rule-spec}.

Additionally, \enclavedom does not require a 1:1 correspondence between
data objects and memory domains.
That is, multiple individual data objects may be placed into the
same domain. Such access rules may be beneficial in cases in which
a number of different in-enclave functions require access to the same
individual data objects.

\subsection{Privilege Isolation for Graphene-SGX}
\label{secsec:graphene-prototype}

As described in~\S\ref{secsec:libos-dev},
libOSes such as Graphene-SGX~\cite{graphene-sgx}
provide application developers with a mechanism for
reaping the benefits of \sgx, while running unmodified applications.
However, Graphene-SGX's trusted shim layer performing
privileged operations (\ie system calls) runs with the same privileges
as the application binary and any third-party libraries imported into the
application.
This lack of privilege isolation poses a threat to the libOS and any
applications it runs because vulnerabilities or attacks
in untrusted third-party code may corrupt or leak sensitive internal
libOS data.

As a preliminary step to demonstrate \enclavedom's ability to
improve the protection of
large-TCB applications against third-party code,
we have built a \enclavedom-enabled prototype of
the Graphene-SGX libOS implementing a privilege isolation mechanism.
Our current prototype enforces privilege isolation by placing all file
system-related data structures into memory domains, and by only granting
access to this data within relevant system calls.
More specifically, we declare two memory domains: \emph{handle\_dom}
contains all file descriptor-specific metadata, and \emph{fs\_dom} contains
all file system management metadata (\eg the mount table).

Our decision to create these two domains was largely based on
Graphene-SGX's implementation of the file system interface,
which treats these two types of metadata separately employing
a separate memory manager for each. As such, we were able to replace
Graphene-SGX's internal memory allocation calls in these memory managers
with \enclavedom domain memory management API calls for the corresponding
memory domains.

We currently declare data access rules for 32 libOS system calls,
for which we automatically generate the \enclavedom policy code
with our policy generator. However, porting \enclavedom to Graphene-SGX
required us to manually implement the sandboxing wrapper functions for
the 32 system calls since the libOS employs a custom interface to \sgx;
this interface is unfortunately not compatible with the official \sgx SDK,
hindering us from automatically generating the \enclavedom system call wrappers
with the Edger8r tool. Furthermore, to comply with Graphene-SGX's system
ABI and make \enclavedom API calls from the Library OS layer,
we needed to create additional wrapper functions for \enclavedom's API in the
Host layer.

Nevertheless, our instrumentation only affects the Graphene-SGX libOS itself,
so application developers who do not wish to provide additional
application-level data protections can still remain agnostic to
\enclavedom (as well as \sgx) by running unmodified applications on top of our
\enclavedom-enabled Graphene-SGX prototype.

\subsection{\enclavedom Code Generation}
\label{secsec:enclavedom-code}

\Paragraph{Policy generator.} \enclavedom
improves the protection of
the developer's policy against tampering by untrusted
code by leveraging \sgx's measurement of static enclave code.
To avoid having developers manually create an internal representation
of their policy, we provide developers with a \emph{policy parser}
that automatically populates
\enclavedom's policy data structures based on the developer's policy, and 
generates a header file containing the internal representation of the
data object ACL. 

Developers run the policy generator as an additional step in their
application build process, and include the generated ACL header file in the
libOS or application source code.
The \enclavedom API can then access these policy data
structures directly at run time.

Finally, the policy generator tool allows libOS and \sgx-native
application developers to configure the
number of enclave pages reserved for a memory domain. If no domain pool size
is specified, the default of four 4-KB pages is used.

\Paragraph{\enclavedom Edger8r tool.}
Since most \sgx-native applications do not compute on static sensitive data,
developers must often pass sensitive data objects between the untrusted
context and the enclave at run time. 
To examine how much we can reduce the burden on developers
and to help to ensure that \enclavedom-protected \sgx-native applications
enhance the protection of sensitive data objects,
we built an experimental \enclavedom-aware Edger8r tool that automatically inserts
data object and domain memory allocation calls
in the generated trampoline code based on the developer's \enclavedom policy.

In addition, \enclavedom can bootstrap the vanilla Edger8r's
functionality to create
trampoline code to create application-specific wrappers around privileged
in-enclave functions to create the function execution sandboxes. These
generated sandboxing wrappers include calls to \enclavedom's execution sandboxing
API surrounding the actual call to the privileged function. Developers would
then only be required to replace the calls to the original function with
calls to the sandboxing wrapper function to ensure that the relevant
sensitive data objects may be accessed within the scope of the sandboxed
function. Any ``un-sandboxed'' calls to the original function would proceed
but attempts to access sensitive data objects will be blocked by \enclavedom's
Monitor.

\Paragraph{\enclavedom for containerized applications.}
While not currently implemented in our Graphene-SGX prototype,
libOS developers may wish to provide support for \enclavedom to
\emph{application} developers. 
To keep the burden on the application developers minimal,
they would only be required to specify a \enclavedom policy specifying
the privileged third-party functions their application calls
and the sensitive data objects these functions may access. After automatically
generating the ACL header file and including this header file into their
application's source code.
We then envision Graphene-SGX transparently parsing
the application's data object ACL header file, allocating the
developer-specified data objects in their respective memory domains,
and creating the application-specific \enclavedom's dynamic execution
sandboxes at run time.

\section{Evaluation}
\label{sec:eval}

We evaluate \griffin's security properties
by performing a case study of an adversarial application that we
run in our prototype, and present an
analysis of additional hypothetical vulnerabilities in enclave code.
To evaluate the performance and memory overheads
\griffin imposes on the Graphene-SGX libOS, we
ran microbenchmarks to understand how
the underlying \griffin operations affect our measurements.

Our tests were performed on a machine with an
experimental Intel CPU with two 224 GB Intel SSDs,
running Ubuntu 17.04 on Linux Kernel 4.10.0-42-generic.
Intel CPUs that feature both \sgx and MPK are unavailable commercially
at the time of writing.\footnote{As of May 2020, we have successfully
  demonstrated \griffin on an Intel Core i7-1065G7 (Ice Lake) platform.}
Further, the policy for our prototype contains access rules for 32
libOS system calls, and is configured
to provision a maximum of four 4-KB memory pages per domain.

\subsection{Security Analysis}
\label{secsec:security-eval}

To evaluate \griffin's security properties, we first do a case study
on an application that attempts to tamper with Graphene-SGX's
file descriptor table. We developed
this adversarial application in our laboratory setting,
demonstrating the feasibility of such attacks without \griffin's
protections, and \griffin's broader applicability to similar attacks.
Furthermore, we also analyze additional hypothetical
vulnerabilities and memory attacks in enclave code.

\Paragraph{Case study: Descriptor Table Corruption.}
To demonstrate \griffin's ability to mitigate unauthorized
accesses, we examine the instance of file descriptor table
corruption in a libOS-based \sgx application.
If successful, such attacks can be especially detrimental since
they may enable a wide range of malicious application behaviors.

In this case study, the adversarial application
imports our specially crafted malicious library which
attempts to corrupt Graphene-SGX's file descriptor table (or FD table).
We ran this adversarial application in vanilla Graphene-SGX and our
\griffin-enabled Graphene-SGX prototype, and found that the attack succeeds
in vanilla Graphene-SGX, where \griffin helps
prevent this attack via a segfault.

This attack is possible in vanilla Graphene-SGX
because the libOS itself and the adversarial
application run in the same address space.
Recall from~\S\ref{secsec:graphene-prototype} that our prototype
places the FD table in the \emph{handle\_dom} memory domain, which is only
accessible during those libOS system calls specified in our prototype's
data access policy. In other words, since our crafted library does not
make system calls that \griffin executes in a sandbox, but rather
attempts to access the FD table via a reverse-engineered
code path that does not run inside a dynamic sandbox, access to the
\emph{handle\_dom} is never granted to the crafted library.

While the attack in this case study may seem contrived at first glance,
we believe that file descriptor table corruption demonstrates
a broader vulnerability class that allows third-party code
to abuse its access to in-enclave data.
\griffin mitigates this threat to TEE applications
which arises specifically because developers include untrusted
third-party code.

In addition, this case study shows that \griffin can help prevent
a more insidious class of attacks capable of circumventing
more traditional access control methods
such as system call interposition: since our studied attack does not
require direct system calls to access a sensitive data object,
interposition mechanisms that verify that the application is only
accessing authorized OS resources would not even be invoked.

\Paragraph{Hypothetical vulnerabilities.} 
We consider a broad range of vulnerabilities that untrusted third-party code
running inside an enclave may hypothetically be able to exploit
in order to leak sensitive in-enclave data to the untrusted application
component.
Table~\ref{tab:hypo-vulns} summarizes three hypothetical vulnerabilities,
and the specific \griffin mechanism that mitigates the vulnerability.

\begin{table*}[t]
  \caption{\label{tab:hypo-vulns} Hypothetical vulnerabilities in
    enclave code, mitigation strategies, and whether
    \griffin currently implements them.}
  \begin{center}
  \begin{tabular}{ccc}
    \toprule
 \textbf{Vuln Type} & \textbf{Mitigation} & \textbf{Implemented?} \\
 \hline
 confused deputy attack & execution sandbox & Y  \\
WRPKRU misuse & static analysis & N \\
MPK OS-level API misuse & syscall interposition & N \\
\bottomrule
  \end{tabular}
  \end{center}
\end{table*}

\griffin allows developers to grant specific in-enclave
functions privileges to access certain sensitive data objects.
However, an untrusted third-party function running inside the enclave
may attempt to escalate its own privileges by calling a more
privileged function in order to gain unauthorized access to
sensitive data objects. \griffin helps
prevent such \emph{confused deputy attacks}~\cite{confused-deputy}
via its dynamic execution sandbox mechanism (see~\S\ref{secsec:monitor}).

Untrusted in-enclave code may attempt to manipulate the value of the
PKRU by including direct \texttt{WRPKRU} instructions within its code.
Since MPK does not currently verify the origin of \texttt{WRPKRU}
instructions in a userspace process, one viable mitigation strategy
that \griffin could employ is static analysis and binary instrumentation
in order to detect and replace errant \texttt{WRPKRU}
instructions in non-\griffin enclave code (similarly as in ERIM~\cite{erim}).

Finally, since MPK relies on the OS for provisioning
protection keys, in-enclave code may attempt to gain authorized access to
MPK-protected memory or corrupt the PKRU by directly making
\texttt{pkey\_mprotect} or \texttt{pkey\_set} system calls via an ocall.
One potential technique for mitigating these issues may be system call
interposition in the kernel to monitor the origin of these calls.

\subsection{Performance Microbenchmarks}
\label{secsec:micros}

\begin{table}[t]
  \caption{\label{tab:syscalls-micro} Mean percentage of execution time
    spent performing \griffin operations for five libOS syscalls. }
  \begin{center}
  \begin{tabular}{ccc}
    \toprule
 & \textbf{\% time in \griffin} & \textbf{accessed memdom(s)} \\
\hline
\texttt{open} & 6.4 & handle, fs \\
\texttt{close} & 49.1 & handle \\
\texttt{stat} & 49.9 & fs \\
\texttt{fstat} & 50.1 & handle, fs \\
\texttt{mmap} & 0.8 & handle \\
\bottomrule
  \end{tabular}
  \end{center}
\end{table}

\begin{table}[t]
  \caption{\label{tab:mem-usage} Peak memory usage for internal \griffin
    data structures for each domain as well as Graphene-SGX as a whole in our
    prototype.}
  \begin{center}
  \begin{tabular}{cc}
    \toprule
 & \textbf{memory usage (in bytes)} \\
\hline
\textit{handle\_dom} & 98 \\
\textit{fs\_dom} & 1030 \\
Total & 1200 \\
\bottomrule
  \end{tabular}
  \end{center}
\end{table}

\Paragraph{Execution time overhead.}
To analyze the impact of \griffin on the performance of
the Graphene-SGX libOS, we took microbenchmarks of
the \texttt{open}, \texttt{stat}, \texttt{fstat}, \texttt{mmap},
and \texttt{close} system calls. Recall from~\S\ref{sec:implementation}
that our \griffin-enabled prototype of Graphene-SGX
isolates the libOS internal file descriptor and file system
management data structures (e.g. mount table) 
into two memory domains. Thus, we chose to benchmark these
five system calls since they require access either only to the
file descriptor domain \textit{handle\_dom}, the file system management 
domain \textit{fs\_dom}, or both, and sought to determine 
how accessing the different memory domains affects performance.

We used the \texttt{lat\_syscall} benchmark of LMbench 2.5~\cite{lmbench},
which stress tests six system calls; for each LMbench experiment, we
then measured the amount of execution time of each our target system calls
spent performing \griffin operations.
Table~\ref{tab:syscalls-micro} shows the mean percentage of
execution time that \griffin operations comprise in our
five benchmarked syscalls 
as well as the \griffin memory domain that each syscall accesses internally.

We observe two groups of system calls: the first group, \texttt{mmap}
and \texttt{open}, spends only a small portion of execution time (at most
6.4\%) performing \griffin operations, while the second group of syscalls
spends about 50\% of the execution time performing \griffin operations.
This result is perhaps not entirely unsurprising given that the primary
purpose of the system calls in the second group is to access the file
system metadata, while \texttt{open} and \texttt{mmap} perform a
much larger number of tasks beyond manipulating the internal file system data
structures. 

\Paragraph{Memory overhead.}
Our evaluation additionally sought to quantify the memory overhead that
\griffin imposes on Graphene-SGX. Specifically, we measure
the additional memory required to maintain the application ACL
and internal memory domain management data structures for each domain
as well as for Graphene-SGX. Table~\ref{tab:mem-usage} shows the median
peak memory consumption in bytes of \griffin's internal data structures
for our Graphene-SGX prototype as a whole as well as each memory domain
during our syscall microbenchmarks above.

We find that \griffin's memory overhead is very modest requiring only
an additional 1.2 KB of memory for the entire Graphene-SGX libOS, and a
mean of 0.6 KB per memory domain. Note that our configuration provisions
far more memory per domain than is required.

\section{Related Work}
\label{sec:related}

\Paragraph{In-\sgx memory protection.}
A small number of prior proposals have sought
to enhance memory protection \emph{within} an
\sgx enclave. SGXBounds~\cite{sgxbounds} uses tagged
pointers for efficient bounds-checking within
an \sgx enclave. SGX-Shield~\cite{sgx-shield} provides
a new ASLR scheme to improve the protection of
\sgx applications against
memory corruption attacks inside the enclave. T-SGX~\cite{tsgx}
leverages the hardware transactional memory provided by \intelreg
TSX in order to help prevent controlled-channel attacks in an \sgx enclave.
Multi-domain SFI~\cite{md-sfi} subdivides an \sgx enclave
into multiple memory domains using MPX's
hardware-enforced memory bounds checking to implement privilege isolation
and multi-process library OSes within a single enclave.

All of these prior proposals (except Multi-domain SFI) aim to
prevent a particular class of memory attacks.
In contrast, \enclavedom operates at the granularity of application-level
data objects to help developers prevent sensitive data leaks in a more
intuitive fashion. Multi-domain SFI is closest to our approach, but we believe
\enclavedom's use of MPK, a technique designed specifically for memory access
control, provides a more adequate solution for our goals.

\Paragraph{Intel hardware memory protection in legacy applications.}
Numerous prior proposals have leveraged novel
techniques available in Intel processors
to help protect sensitive application data
within a single address space. 

ERIM~\cite{erim}, MemSentry-PKU~\cite{memsentry} and
Janus~\cite{janus-pku} use MPK to partition a legacy application's 
address space into two or more domains to isolate sensitive application data.
\enclavedom builds on this prior research generalizing the use of MPK
to implement multiple
memory domains and applies this mechanism to
\sgx applications.
leveraging MPK to partition an
\sgx enclave and isolate sensitive in-enclave data objects in separate domains.

Systems such as Dune~\cite{dune}, MemSentry~\cite{memsentry},
and Janus~\cite{janus-pku} rely on \intelreg VT-x virtualization hardware
to create intra-process isolated compartments. Unlike \enclavedom,
the main goal of these approaches is to provide \emph{execution}
isolation for running different application components in separate
address spaces. \enclavedom's main goal is to help
protect sensitive
application data shared between different application components
by controlling access to sensitive data objects at run time.

\Paragraph{Non-Intel hardware memory isolation.}
Other architectures, most notably ARM, also provide
techniques designed for finer-grained memory access control.
For instance, ARMlock~\cite{armlock}, Shreds~\cite{shreds},
and FlexDroid~\cite{flexdroid} leverage ARM memory domains, a technique
very similar to MPK, to allow developers to create a number of
isolated execution regions with associated private memory compartments. 

\Paragraph{Process isolation.} 
A large number of prior proposals have
leveraged OS primitives to restrict
application components in legacy and mobile applications.
Systems such as Wedge~\cite{wedge}, Privtrans~\cite{privtrans},
Passe~\cite{passe}, CodeJail~\cite{codejail}, AdSplit~\cite{adsplit}
and BreakApp~\cite{breakapp} partition an
application into multiple processes to run unprivileged components
in isolated address spaces.
Ryoan~\cite{ryoan} partitions an \sgx
application into separate \sgx processes
as a means to isolate different application components.

While such process-based isolation provides strong memory protections,
these approaches require significant development efforts to re-architect
a monolithic application for a multi-process model. \enclavedom's
single-process design, on the other hand, provides a more practical approach
for \sgx application developers that does not require major application
refactoring, and still provides strong hardware-assisted memory isolation.

\Paragraph{Intra-process isolation.}
Arbiter~\cite{arbiter}, SMV~\cite{smv},
Light-weight Contexts~\cite{lwc}, and Pyronia~\cite{pyronia}
partition a single process address space using multiple page tables
in order to control access to different memory compartments at
intra-process granularities (\eg thread-level in Arbiter or SMV).
\enclavedom borrows many concepts from these
intra-process isolation proposals, but relies on more efficient hardware-based
techniques that are more adequate for use in \sgx applications.

\section{Conclusion}
\label{sec:conclusion}

We have presented \enclavedom, a privilege separation system
for large-TCB applications running in trusted execution environments.
\enclavedom combines the security of TEEs with memory tagging to help
prevent sensitive data leaks and corruption by untrusted third-party
code imported into containerized and TEE-native applications.
To avoid partitioning an enclave into multiple TEE processes,
\enclavedom isolates sensitive data objects
within a single enclave address space. As such, \enclavedom
supports data sharing between different in-enclave functions,
while enforcing least privilege data access policies.

We have implemented \enclavedom as a userspace API for
large-TCB \sgx applications that uses MPK for enclave memory tagging.
To demonstrate \enclavedom's properties, we implement OS privilege isolation
in the Graphene-SGX library OS. Our evaluation of our \enclavedom-enabled
Graphene-SGX prototype imposes a very modest memory overhead and can help
protect containerized applications
against unauthorized accesses to sensitive in-enclave data.

\section*{Acknowledgments}
We thank Mingwei Zhang, Michael Steiner, 
Bruno Vavala, Prakash Narayana Moorthy,
Dmitrii Kuvaiskii, Mona Vij, Michael LeMay, Thomas Knauth,
and Vinnie Scarlata for
their feedback and insightful discussions.

{\normalsize \bibliographystyle{acm}
\bibliography{references}}

\end{document}